\documentclass[manuscript]{aastex62}
\usepackage{amsmath,amssymb}
\usepackage{graphicx}
\usepackage{hyperref}

\begin{document}

\title{Spectroscopic Observations of High-speed Downflows in a C1.7 Solar Flare}
\author{Yi-An Zhou}
\affiliation{School of Astronomy and Space Science, Nanjing University, Nanjing 210023, People's Republic of China}
\affiliation{Key Laboratory for Modern Astronomy and Astrophysics (Nanjing University), Ministry of Education, Nanjing 210023, People's Republic of China}
\author{Y. Li}
\affiliation{Key Laboratory of Dark Matter and Space Astronomy, Purple Mountain Observatory, Chinese Academy of Sciences, Nanjing 210033, People's Republic of China}
\author{M. D. Ding}
\affiliation{School of Astronomy and Space Science, Nanjing University, Nanjing 210023, People's Republic of China}
\affiliation{Key Laboratory for Modern Astronomy and Astrophysics (Nanjing University), Ministry of Education, Nanjing 210023, People's Republic of China}
\author{Jie Hong}
\affiliation{School of Astronomy and Space Science, Nanjing University, Nanjing 210023, People's Republic of China}
\affiliation{Key Laboratory for Modern Astronomy and Astrophysics (Nanjing University), Ministry of Education, Nanjing 210023, People's Republic of China}
\author{Ke Yu}
\affiliation{School of Astronomy and Space Science, Nanjing University, Nanjing 210023, People's Republic of China}
\affiliation{Key Laboratory for Modern Astronomy and Astrophysics (Nanjing University), Ministry of Education, Nanjing 210023, People's Republic of China}

\correspondingauthor{M. D. Ding}
\email{dmd@nju.edu.cn}

\begin{abstract}

In this paper, we analyze the high-resolution UV spectra for a C1.7 solar flare (SOL2017-09-09T06:51) observed by the \textit{Interface Region Imaging Spectrograph} (\textit{IRIS}).  {We focus on the spectroscopic observations at the locations where the cool lines of \ion{Si}{4} 1402.8 \AA\ ($\sim$10$^{4.8}$ K) and \ion{C}{2} 1334.5/1335.7 \AA\ ($\sim$10$^{4.4}$ K) reveal significant redshifts with Doppler velocities up to $\sim$150 km s$^{-1}$.} These redshifts appear in the rise phase of the flare, then increase rapidly, reach the maximum in a few minutes, and proceed into the decay phase. Combining the images from \textit{IRIS} and Atmospheric Imaging Assembly (AIA) on board the {\em Solar Dynamics Observatory} ({\em SDO}), we propose that the redshifts in the cool lines are caused by the downflows in the transition region and upper chromospheric layers, which likely result from a magnetic reconnection leading to the flare. In addition, the cool \ion{Si}{4} and \ion{C}{2} lines show gentle redshifts (a few tens of km s$^{-1}$) at some other locations, which manifest some distinct features from the above locations. This is supposed to originate from a different physical process.
\end{abstract}


\section{Introduction}
\label{sec-intro}

Solar flares are one of the most energetic events on the Sun (e.g., \citealt{Fletcher2011}), which are generally believed to be associated with magnetic reconnection (\citealt{Kopp1976, Masuda1994, Lin2000}). In the standard flare model, magnetic reconnection releases massive energy in the corona, which is pre-stored in a non-potential magnetic structure. The released energy is subsequently transported downward to the lower atmosphere through thermal conduction and/or nonthermal particles. Hence the chromospheric plasma is heated and emits strong radiation that forms flare ribbons. Due to an enhanced thermal pressure, the heated chromospheric material moves upward to the corona and fills the flare loops that are visible in EUV and soft X-ray bands. This process is known as chromospheric evaporation \citep{Brosius2004, Milligan2006a, Milligan2006b, Doschek2013}. In general, the evaporation is also accompanied by a compression of chromospheric plasma based on momentum balance \citep{Canfield1990}, which is referred to as chromospheric condensation \citep{Fisher1985, Milligan2006a, Zhang2016}.

There are several ways to investigate the energetics and dynamics of flares, of which the spectroscopic diagnostics  {are} a classical and important one. Based on the fact that each spectral line is formed in a specific atmospheric layer, we could obtain various information on different layers of the atmosphere by using different lines. For example, the cool lines of \ion{Si}{4} and \ion{C}{2} that are formed at $\sim$10$^{4.8}$ K and $\sim$10$^{4.4}$ K respectively could be used to diagnose the transition region (TR) and upper chromosphere \citep[e.g.,][]{McIntosh2009, Tian2014a, Li2019}. The hot \ion{Fe}{21} line with a formation temperature of $\sim$10$^{7.1}$ K could reveal the physical properties of the hot corona \citep{Tian2014b, Battaglia2015, grah15, Li2015, Polito2015, Polito2016, Young15, Dudik2016, Brosius2018}. In particular, the Doppler velocity can be derived from line profiles, which is a good indicator of the plasma flows during a flare. For optically thin lines, blueshifts are generally due to plasma upflows whereas redshifts imply plasma downflows.
 
There are a large number of studies on the blueshifts/redshifts which result from chromospheric evaporation/condensation in solar flares. For instance, blueshifts with velocities ranging from $\sim$50 to $\sim$300 km s$^{-1}$ were typically observed in the spectra of highly ionized Fe atoms (e.g., \ion{Fe}{16} to \ion{Fe}{24}) from instruments such as Hinode/EIS and the {\em Interface Region Imaging Spectrograph} ({\em IRIS}) (e.g., \citealt{Brosius2004, Liu2006, Chen2010, Zhang2016, LiD2017}). Redshifts with velocities of $\sim$20--80 km s$^{-1}$ from the relatively cool lines of \ion{He}{2}, \ion{O}{3}, \ion{O}{5}, and \ion{Fe}{12} were also detected owing to plasma condensation (e.g., \citealt{Wuelser1994, Ding1995, Czaykowska1999, Brosius2003, Kamio2005, Teriaca2006, Del2008, Milligan2009}). Note that these blueshifts/redshifts are located on the flare ribbons.

It is worth mentioning that there are few observations that have reported the rapid redshifts (say, $>$100 km s$^{-1}$) in the cool lines on the flare ribbons. Instead, some rapid blueshifts or redshifts were observed on the loops, which could be interpreted as magnetic reconnection outflows. \cite{Sadykov2015} reported a strong jet-like flow with a redshift velocity of $\sim$100 km s$^{-1}$ in the chromospheric \ion{C}{2} and \ion{Mg}{2} lines just prior to a flare, which perhaps comes from the magnetic reconnection region. \cite{Reeves2015} detected intermittent fast downflows ($\sim$200 km s$^{-1}$) in the \ion{Si}{4} line as evidence for magnetic reconnection between the prominence magnetic fields and the overlying coronal fields. Moreover, bidirectional outflows with velocities of tens to hundreds of km s$^{-1}$ were observed in the \ion{Si}{4} line in terms of a tether-cutting (TC) reconnection \citep{Chen2016} or a separator reconnection \citep{Li2017}. In spite of these, spectroscopic observations of magnetic reconnection are still lacking and the detailed physical mechanisms of fast flows appearing in some events are not fully understood yet. 

Fortunately, {\em IRIS} \citep{De14} provides high-resolution slit-jaw images (SJIs) as well as spectra for a large number of flares. The sub-arcsecond observations from {\em IRIS} reveal fine structures of flares and illustrate distinct features of plasma flows. In this work, we detect rapid redshifts with a velocity of $\sim$150 km s$^{-1}$ in the cool \ion{Si}{4} and \ion{C}{2} lines at some locations during a C1.7 flare. We also observe gentle redshifts of tens of km s$^{-1}$ in these cool lines at some other locations. Such distinct redshifts are supposed to originate from different processes. In the following, we present the observations in Section \ref{sec-obs} and data reduction in Section \ref{sec-data}. Then we show the results from {\em IRIS} spectral lines in detail in Section \ref{sec3}. In Sections \ref{sec4} and \ref{sec5}, we give the discussions and conclusions, respectively.

\section{Observations}
\label{sec-obs}

\begin{figure*}
	\centering
	\includegraphics[width=\linewidth]{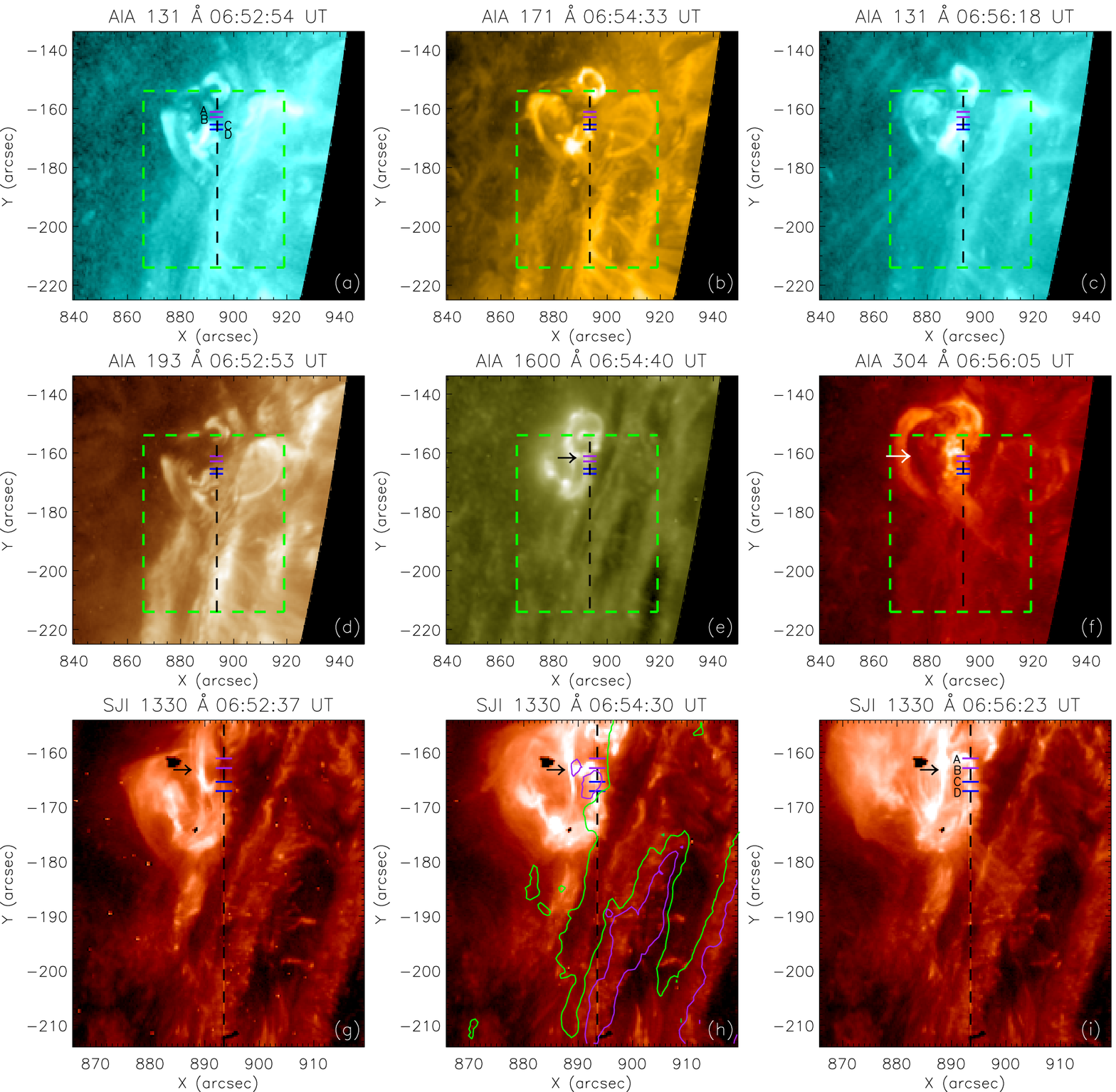}
	\caption{
Overview of the evolution of the C1.7 solar flare in AR 12673. Panels (a)--(f) are for AIA UV and EUV images while panels (g)--(i) are for {\em IRIS} SJIs. In each panel, the two purple bars (marked by A and B) and two blue bars (marked by C and D) represent four locations that are selected for study in the work. In panels (a)--(f), the green dashed box refers to the FOV of the {\em IRIS} SJIs as shown in panels (g)--(i) and the black dashed line marks the position of the {\em IRIS} slit. The black arrow in panels (e) and (g)--(i) denotes the sigmoid structure. In panel (f), the white arrow indicates the filament plasma draining. The purple and green contours in panel (h) indicate the positive and negative magnetic fields observed from HMI, respectively.
}
    \label{fig1}
\end{figure*}

The C1.7 flare under study occurred in the active region NOAA 12673 near the west limb. It started at $\sim$06:51 UT and peaked at $\sim$06:56 UT on 2017 September 9. Figure \ref{fig1} gives an overview of the flare from the Atmospheric Imaging Assembly (AIA; \citealt{Lemen12}) images (Figures \ref{fig1}{(a)--(f)}) as well as {\em IRIS} SJIs (Figures \ref{fig1}{(g)--(i)}) at different passbands. It is seen that a sigmoid structure (marked by the black arrow) shows up in some low-temperature channels, e.g., SJIs at 1330 \AA\ (Figures \ref{fig1}(g)--(i)) and AIA 1600 \AA\ images (Figure \ref{fig1}(e)). Some bright loops or loop-like structures can also be seen in the images. At an early time ($\sim$06:52 UT), a few small flare loops appear at Y $\sim-$168\arcsec\ ({around the  {\em IRIS} slit locations C and D}; see the left panels of Figure \ref{fig1}). At later times ($\sim$06:54 UT and $\sim$06:56 UT), some smaller loop-like structures show up at Y $\sim-$162\arcsec\ (around the slit locations A and B, see the middle and right panels of Figure \ref{fig1} and also the accompanying animation). In Figure \ref{fig1}(h), we overplot the contours of line-of-sight magnetic field observed by the Helioseismic and Magnetic Imager (HMI; \citealt{Sch2012}). One can see that the four slit locations A--D are close to the magnetic inversion region. Note that some jet structures accompanying the flare can be seen in the {\em IRIS} FOV (at Y $\sim-$180\arcsec; see Figures \ref{fig1}(f) and (i)). We also notice that there exists some filament plasma draining in the FOV of {\em IRIS}, which firstly appears at $\sim$06:56 UT (see Figure \ref{fig1}{(f)}, marked by the white arrow) and lasts for $\sim$15 minutes.

\begin{figure*}
	\centering
	\includegraphics[width=\linewidth]{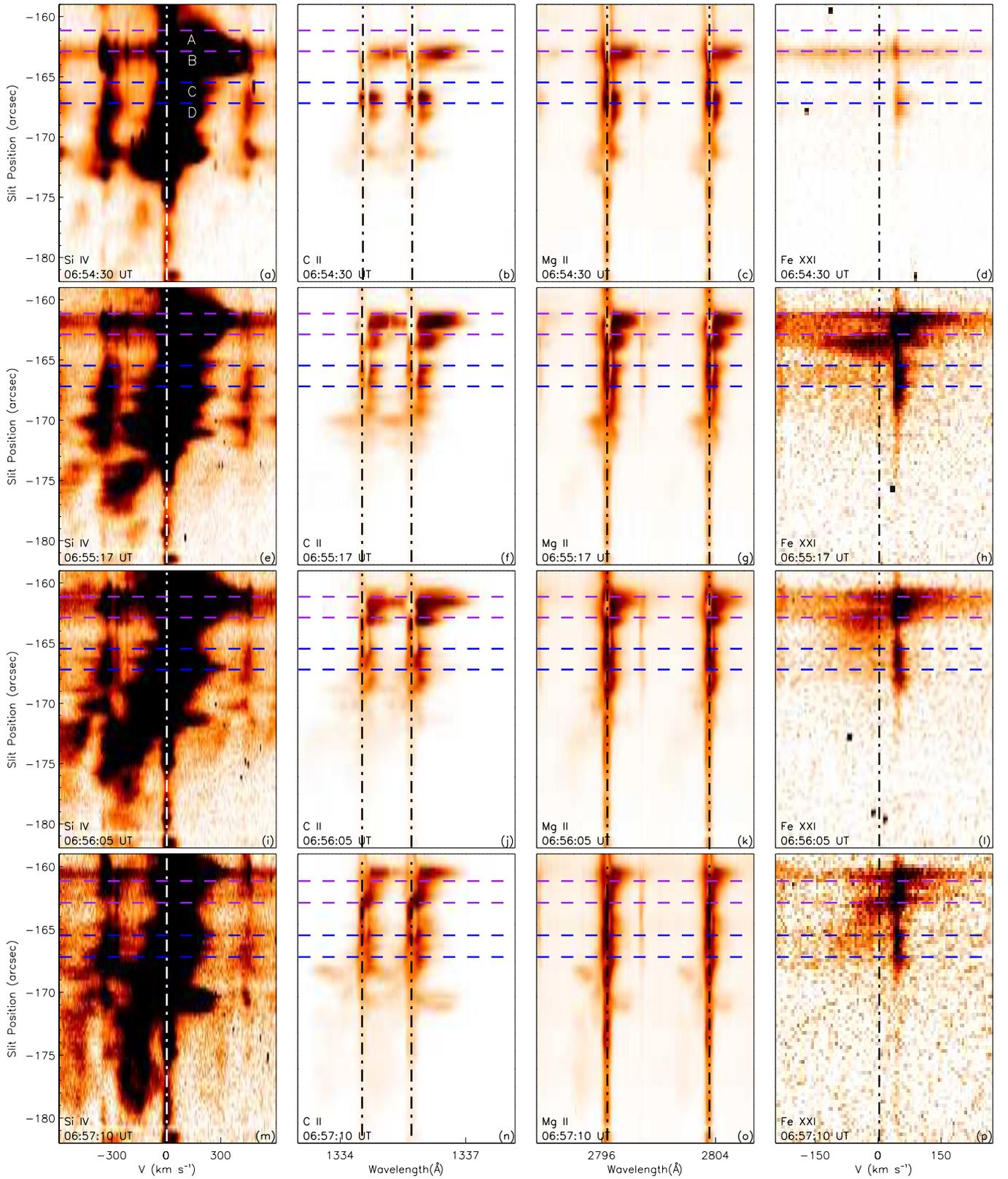}
	\caption{
{\em IRIS} spectra of the Si IV, C II, Mg II, and Fe XXI lines at four selected times during the flare. The vertical dash-dotted lines in each panel refer to the reference centers for each of the lines. The four horizontal dashed lines (purple and blue) refer to the four locations (A--D) that are selected for study. Note that the spectra of the \ion{Si}{4} and \ion{Fe}{21} lines are saturated to show the continuum emission.
}
    \label{fig-spectra}
\end{figure*}

Figure \ref{fig-spectra} shows the spectra of \ion{Si}{4}, \ion{C}{2}, \ion{Mg}{2}, and \ion{Fe}{21} along the {\em IRIS} slit at four times. It is seen that the cool \ion{Si}{4}, \ion{C}{2}, and \ion{Mg}{2} lines show evident redshifts at locations A--D during the flare. Note that there also show up {strong} continuum emission and narrow cool lines {at locations A and B} (see the first column). In addition, one can see some evident emission of the hot \ion{Fe}{21} line at locations A and B, which shows clear redshifts at some times (say, 06:55:17 UT and 06:56:06 UT for location A, also see the line profiles in Figures \ref{fig5}(e) and (g)) except for some blueshifts at location B in part of the time. By contrast, the \ion{Fe}{21} line at locations C and D are much weaker but exhibit evident blueshifts (see Figure \ref{fig-spectra}(p) as well as Figures \ref{fig5}(f) and (h)). In this work, we select these four locations to study the typical features of the flare region, which can be divided into two groups, i.e., one for locations A and B and the other for locations C and D. These two sets of locations exhibit some distinct spectral features on the moment maps as well as in the line profiles as described in Section \ref{sec3}.

\section{Data Reduction}
\label{sec-data}

The C1.7 flare was observed by  {\em IRIS} and the \textit{Solar Dynamics Observatory} (\textit{SDO}). The {\em IRIS} slit was used to preform a medium sit-and-stare spectral observation with a high time cadence of 9.8 s. The slit has a width of 0.\arcsec33 and the pixel scale along the slit is 0.\arcsec166. The spectral resolutions are 0.051 \AA\ and 0.025 \AA\ for NUV and FUV spectra, respectively. {\em IRIS} SJIs at 1330 \AA, 2796 \AA, and 2832 \AA\ have a field of view (FOV) of 60\arcsec $\times$ 62\arcsec\ and a cadence of 28 s. The former two passbands are characteristic of the upper chromosphere and lower transition region \citep{De14}.  {The AIA on board \textit{SDO}} obtained UV and EUV images for this flare with a spatial resolution of 1.\arcsec2 (or 0.\arcsec6 pixel$^{-1}$) and temporal resolutions of 24 s and 12 s, respectively. The UV and EUV bands are sensitive to plasmas at different temperatures. For example, the 131 \AA, 193 \AA, and 171 \AA\ bands are for coronal plasmas with their responses peaking at $\sim$10 MK, $\sim$1.6 MK, and $\sim$0.6 MK, respectively, while the 304 \AA\ and 1600 \AA\ bands are for chromospheric and TR plasmas with formation temperatures of $\sim$0.05 MK and $\sim$0.1 MK, respectively. Note that here, we use the AIA 1700 \AA\ images and SJIs 2832 \AA\ to make a co-alignment, both of which show clear sunspot features. The uncertainty of the co-alignment is estimated to be $\sim$1\arcsec.

We mainly use the \ion{Si}{4} 1402.8 \AA\ line that has a formation temperature of $\sim$10$^{4.8}$ K in this work. The \ion{C}{2} lines at 1334.5 \AA\ and 1335.7 \AA\ ($\sim$10$^{4.4}$ K), the \ion{Mg}{2} k line at 2796.4 \AA\ ($\sim$10$^{4.0}$ K), and the \ion{Fe}{21} line at 1354.1 \AA\ ($\sim$10$^{7.1}$ K) are referred to as well. For the \ion{Si}{4} line, we make a moment analysis to derive the total intensity (the zeroth moment) and Doppler velocity (the first moment) for the following two reasons. (1) The \ion{Si}{4} line profiles are very complicated in this flare, i.e., showing two or even more emission peaks (see Figure \ref{fig4}). (2) The ratio of the two \ion{Si}{4} lines at 1393.8 \AA\ and 1402.8 \AA\ somewhat deviates from 2 (ranging from $\sim$1.4--2.1 at locations A--D) during the flare, indicating that the \ion{Si}{4} line suffers from an opacity effect \citep[e.g.,][]{Peter2014, Yan2015, Kerr2019}. For the optically thin \ion{Fe}{21} line that is mainly contaminated by the \ion{C}{1} 1354.3 \AA\ line, we use a double Gaussian function to fit the two lines. Note that we also apply a triplet Gaussian fitting to these two lines when the \ion{C}{1} line exhibits a red asymmetry during the flare. In order to calculate the Doppler velocity, here we use some photospheric or chromospheric lines over a relatively quiet region before the flare onset to determine the reference wavelength. For the \ion{Si}{4} line, the \ion{S}{1} 1401.5 \AA\ line is used for calibration. For the \ion{Fe}{21} line, the \ion{O}{1} line at 1355.6 Å and the \ion{C}{1} line at 1355.8 \AA\ are used. The reference centriods of the \ion{Si}{4} and \ion{Fe}{21} lines are obtained to be 1402.77 \AA\ and 1354.08 \AA, respectively. {The uncertainty in the Doppler velocity is thus estimated to be $\sim$1 km s$^{-1}$, which is consistent with the velocity accuracy as reported in \cite{De14} and \cite{Wuelser2018}.}

\section{Results}
\label{sec3}

\subsection{{Line profiles and Moment maps}}
\label{sec3.1}

\begin{figure*}
	\centering
	\includegraphics[width=0.9\linewidth]{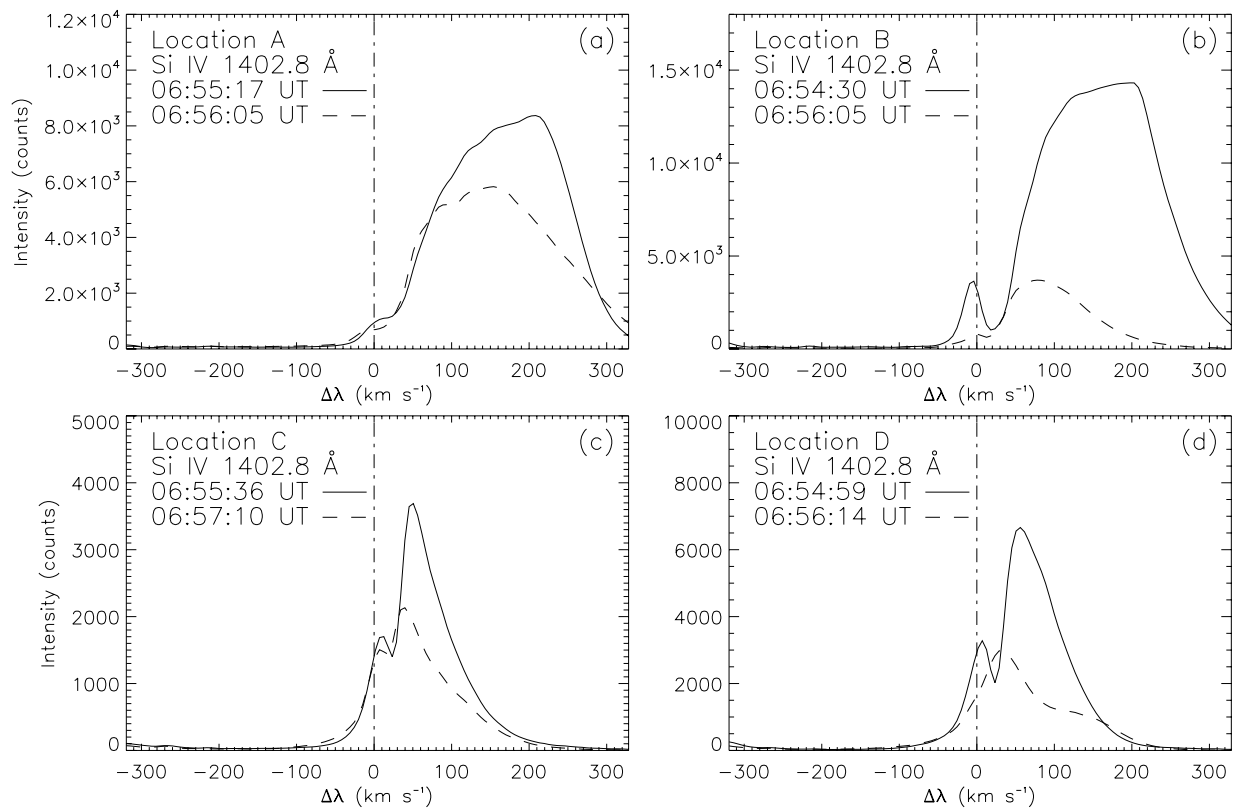}
	\caption{
Some typical profiles of the \ion{Si}{4} 1402.8 \AA\ line at the four locations A--D. In each panels, the vertical dash-dotted line refers to the reference line center. The solid and dashed curves represent the \ion{Si}{4} line profiles at two different times for each of the locations, which are marked by the vertical dash-dotted lines in Figures \ref{fig6} and \ref{fig7}.
 }
    \label{fig4}
\end{figure*}

\begin{figure*}
	\centering
	\includegraphics[width=0.85\linewidth]{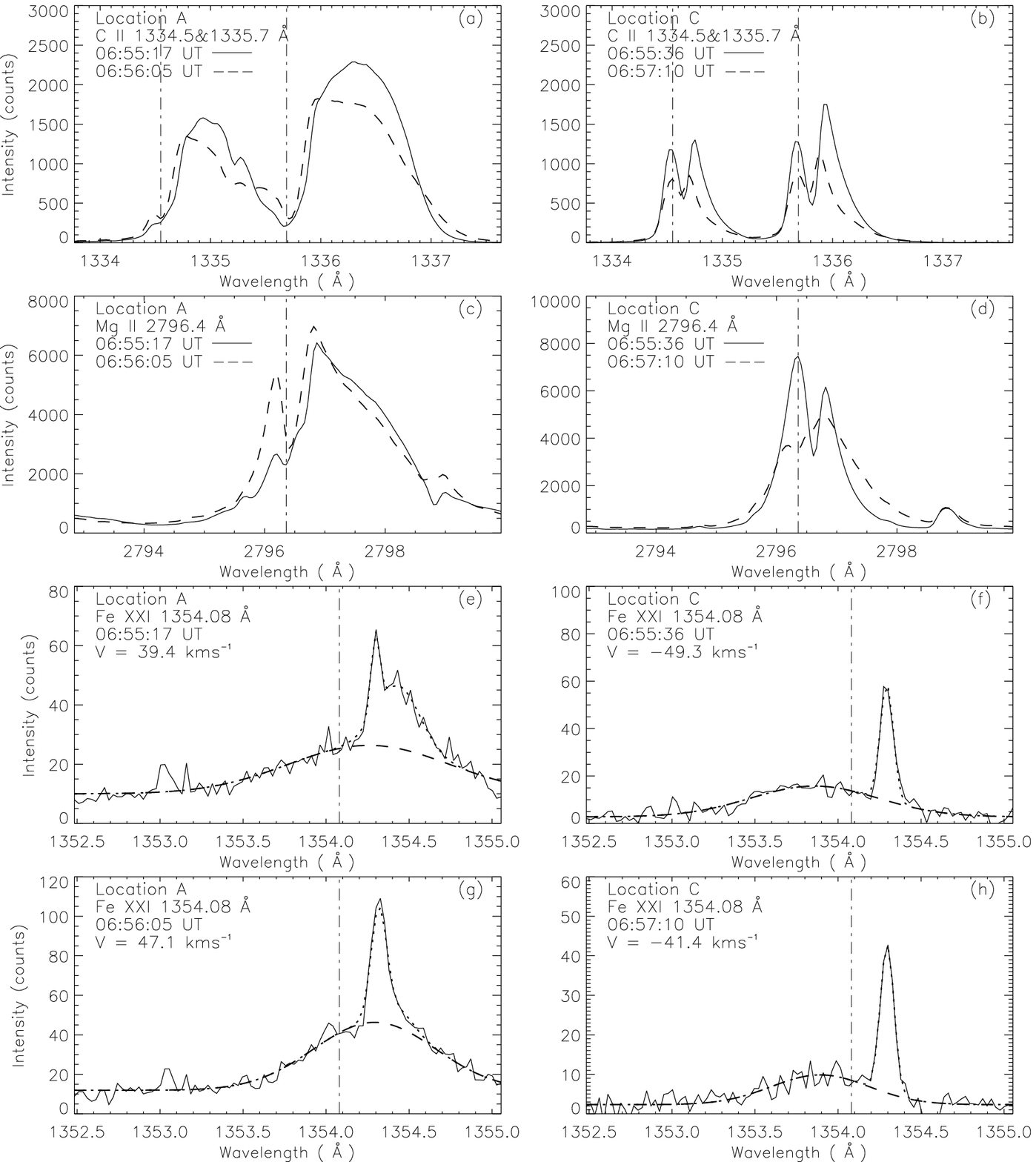}
	\caption{
Line profiles of \ion{C}{2} 1334.5 \AA\ and 1335.7 \AA, \ion{Mg}{2} 2796.4 \AA\ and \ion{Fe}{21} 1354.1 \AA\ at locations A and C. The vertical dash-dotted line in each panel refers to the reference center of each line. In panels (a)--(d), the solid and dashed curves represent the observed line profiles at two different times that are marked in Figures \ref{fig6} and \ref{fig7}. In panels (e)--(h), the observed line profiles are plotted in solid curves, while the fitted line profiles and the \ion{Fe}{21} 1354.1 \AA\ components are plotted in dotted and dashed curves, respectively. Note that in panels (e) and (g), the observed line profile is fitted by a triplet Gaussian function when the \ion{C}{1} line shows a red asymmetry. The positive and negative velocities labelled in panels (e)--(h) refer to the redshift and blueshift velocities from the \ion{Fe}{21} 1354.1 \AA\ line, respectively.
}
    \label{fig5}
\end{figure*}

Figures \ref{fig4} and \ref{fig5} show the line profiles at the four selected locations for some noticeable times when the Doppler velocity of \ion{Si}{4} shows a peak or sub-peak, as indicated by the vertical dash-dotted lines in Figures \ref{fig6} and \ref{fig7}. It is seen that at locations A and B, the \ion{Si}{4} line are mostly redshifted with the Doppler velocity being over 100 km s$^{-1}$ (Figures \ref{fig4}(a) and (b)). For the cooler \ion{C}{2} and \ion{Mg}{2} lines at location A (and also at location B), they show a much stronger red-peak emission with {an unshifted line core (or a central reversal; see Figures \ref{fig5}(a) and (c))}. Note that the two \ion{C}{2} resonance lines are blended due to a large redshift. By contrast, at locations C and D, the \ion{Si}{4} line profiles show two components or a red asymmetry with the Doppler velocity lower than 100 km s$^{-1}$ ({Figures} \ref{fig4}(c) and (d)). In particular, the \ion{C}{2} and \ion{Mg}{2} lines exhibit a redshifted line core at these two locations (Figures \ref{fig5}(b) and (d)). Moreover, weak but blueshifted emission is detected in the hot \ion{Fe}{21} line at location C (Figures \ref{fig5}(f) and (h)). However, at location A, the \ion{Fe}{21} line emission is relatively stronger as well as show evident redshifts with a velocity of a few tens of km s$^{-1}$ (Figures \ref{fig5}(e) and (g)).

\begin{figure*}
	\centering
	\includegraphics[width=0.9\linewidth]{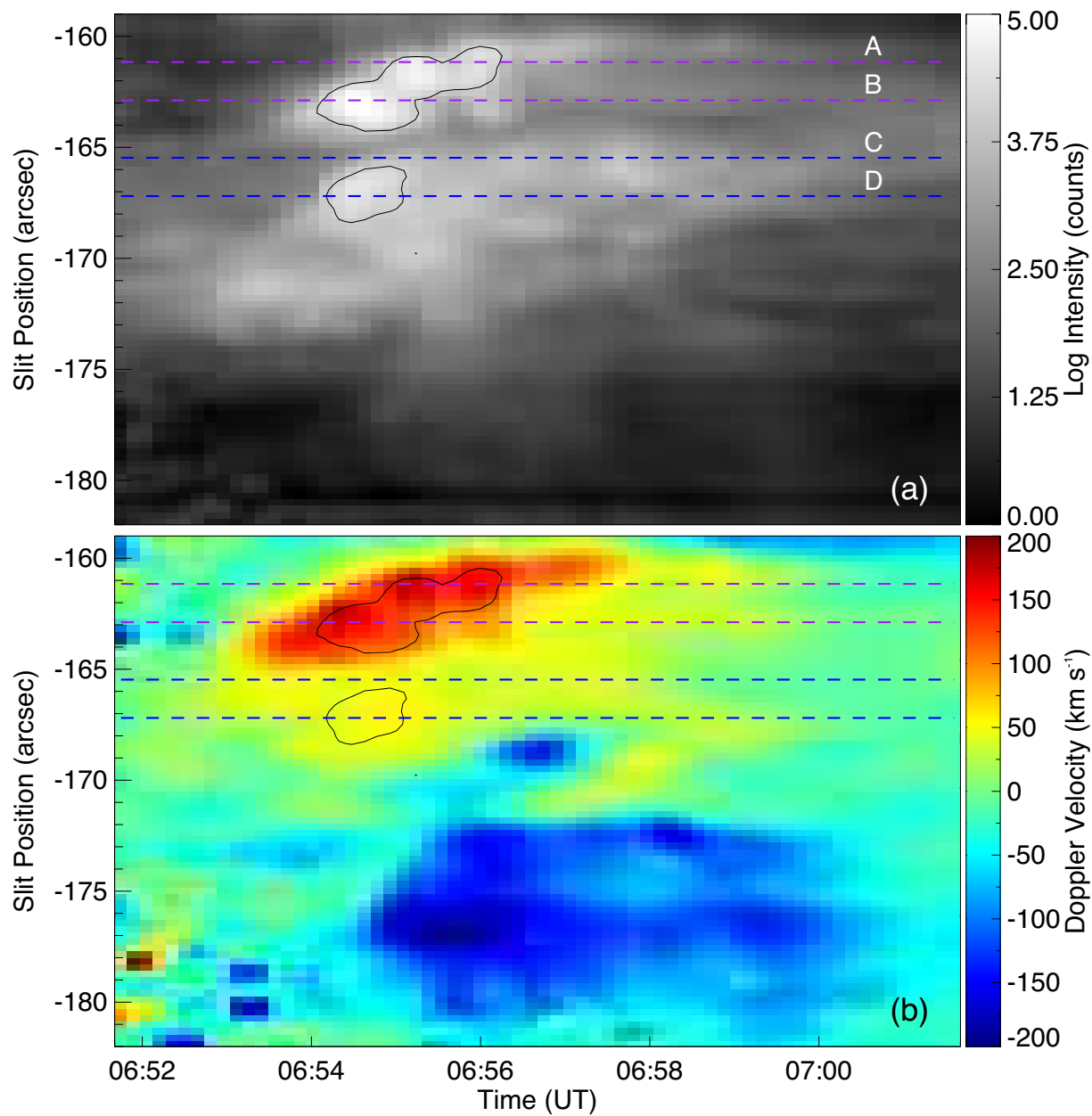}
	\caption{
Space-time diagrams of the \ion{Si}{4} line intensity (a) and Doppler velocity (b) derived from the moment method.  {The four horizontal lines (purple and blue) in each panel refer to the four locations (A--D) that are selected for study.} The contours mark an intensity level of $\log\ $I = 4.2 counts for the \ion{Si}{4} line. In panel (b), the positive and negative {velocities refer to the redshifts and blueshifts}, respectively.
}
    \label{fig3}
\end{figure*}

Figure \ref{fig3}(a) gives the space-time diagram of the \ion{Si}{4} line intensity. One can see that locations A--D show significant brightenings that exhibit an apparent motion towards the north over time. From the Doppler velocity map of the \ion{Si}{4} line in Figure \ref{fig3}(b), it is seen that all these locations display evident redshift features that correspond to the brightenings (see the overplotted contours). However, the redshift velocities at locations A and B can be as high as 150 km s$^{-1}$, which seem to be uncommon in observations. By contrast, the redshift velocities at locations C and D are only a few tens of km s$^{-1}$, which are often observed and reported in previous studies. Note that there appear some notable blueshifts at the bottom part, which are supposed to be caused by the flare-accompanied jet eruptions. 

\subsection{Kinematic Features of the Downflows}
\label{sec3.2}

\begin{figure*}
	\centering
	\includegraphics[width=0.9\linewidth]{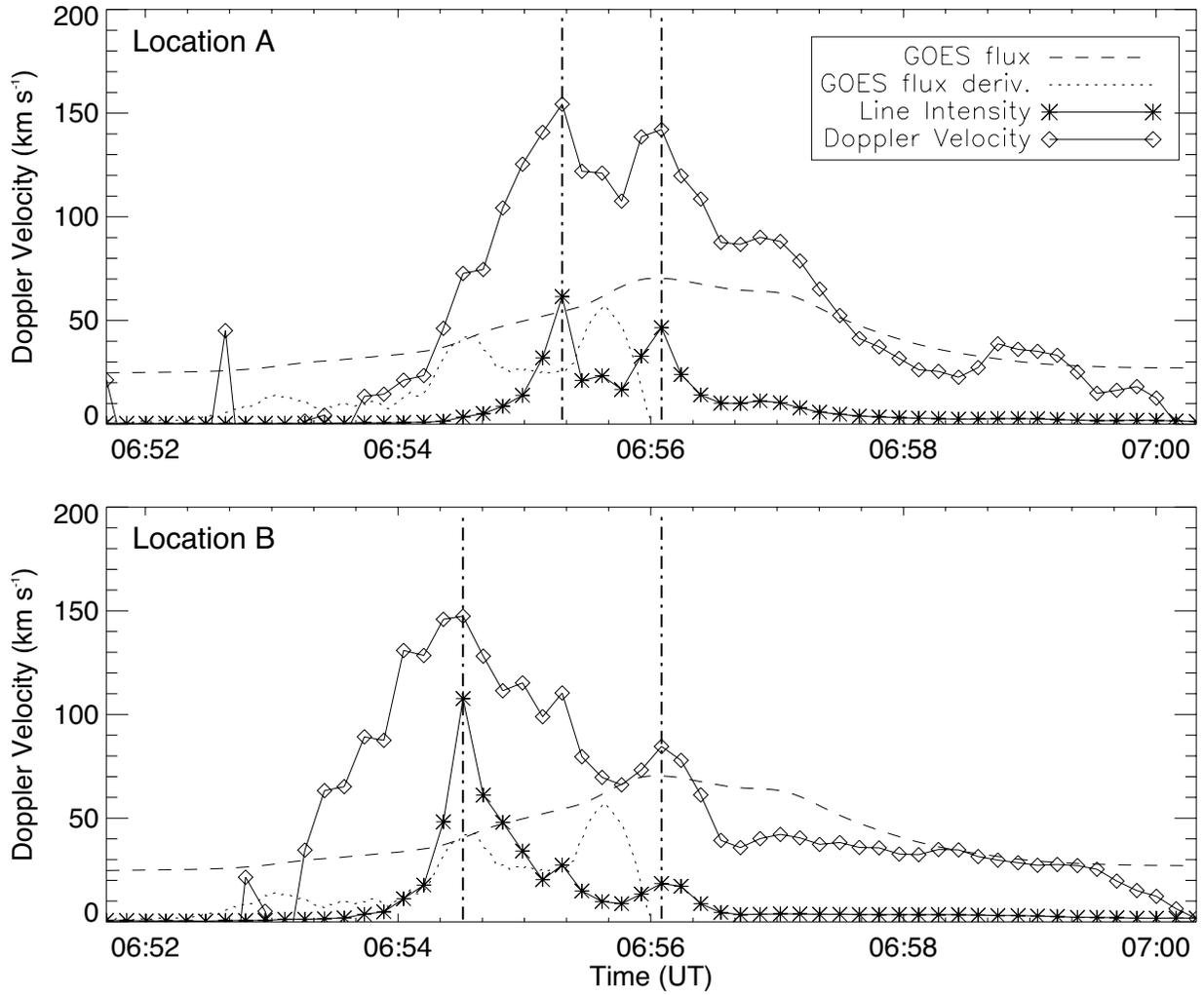}
	\caption{Time evolutions of the Doppler velocity and the wavelength-integrated intensity of the \ion{Si}{4} line at locations A and B.  {The dashed and dotted lines in each panel denote the {\em GOES} SXR flux and its time derivative, respectively.  
The solid line marked with stars shows the \ion{Si}{4} line intensity integrated over wavelength and the solid line marked with diamonds shows the Doppler velocity.} The vertical dash-dotted lines show the peak times of the Doppler velocity. Note that the SXR flux,  {its time derivative, and the \ion{Si}{4} intensity} are in arbitrary units. 
	}
    \label{fig6}
\end{figure*}

\begin{figure*}
	\centering
	\includegraphics[width=0.9\linewidth]{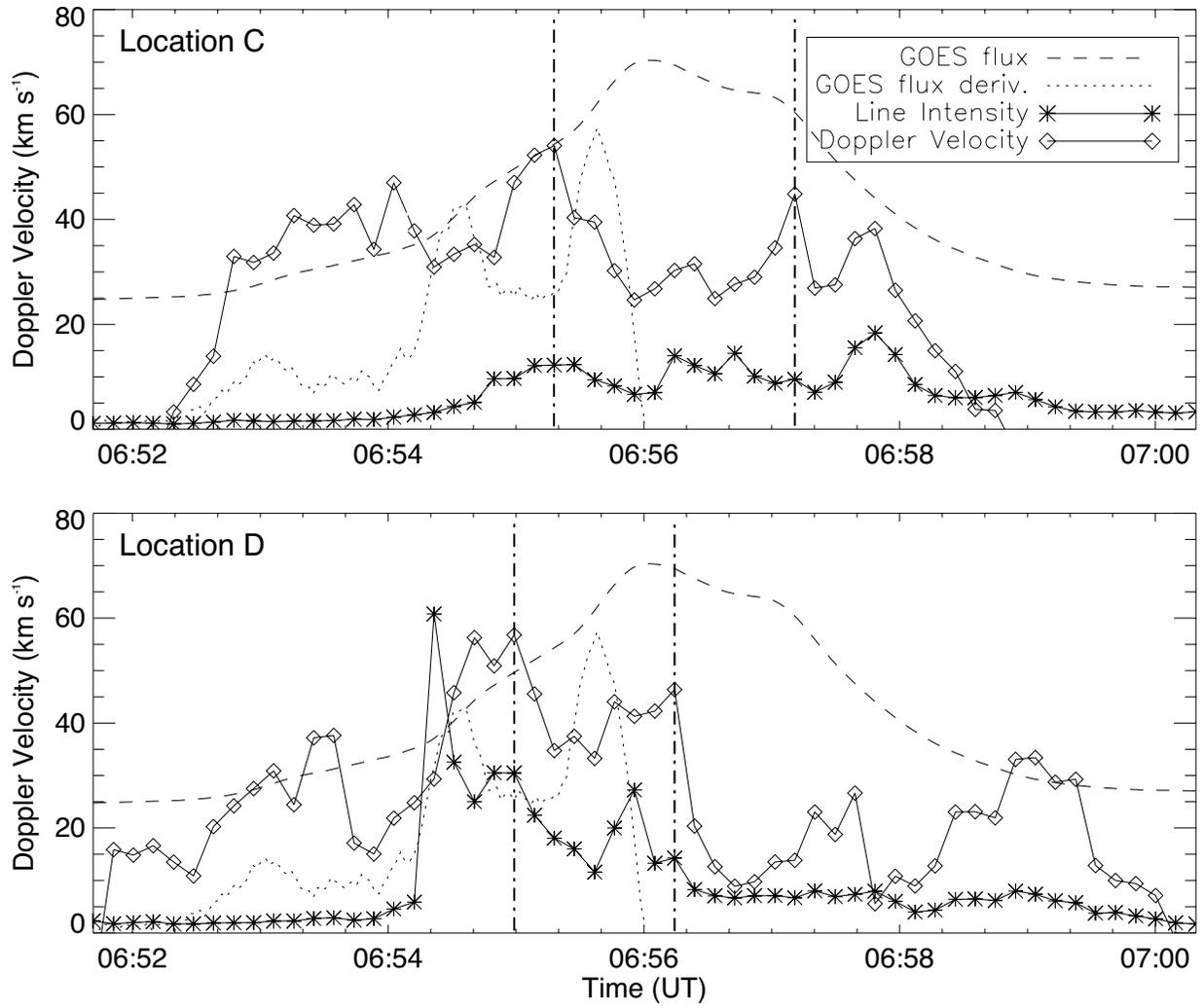}
	\caption{Same as Figure \ref{fig6}, but for locations C and D.}
    \label{fig7}
\end{figure*}

Figures \ref{fig6} and \ref{fig7} show the temporal evolutions of the total intensity (the solid line marked with stars) as well as the Doppler velocity (the solid line marked with diamonds) of the \ion{Si}{4} line at the four locations. Here we also plot the {\em GOES} 1--8 \AA\ soft X-ray (SXR) emission (the dashed line) and its time derivative (the dotted line) to show the evolution of the whole flare. It is seen that the time derivative of SXR emission exhibits two peaks, one at $\sim$06:54:30 UT and the other at $\sim$06:55:40 UT, which may indicate two episodes of energy release.

As shown in the top panel of Figure \ref{fig6}, at location A, the \ion{Si}{4} line intensity starts to rise at $\sim$06:54:20 UT. It shows two peaks with the first one (at $\sim$06:55:17 UT) prior to the main peak of the time derivative of {\em GOES} SXR emission, and the second one (at $\sim$06:56:05 UT) slightly after that (see the two vertical dash-dotted lines). The temporal variation of the redshift velocity resembles that of the line intensity, only that the velocity increases a little bit earlier than the intensity. The velocity rises rapidly from $\sim$06:53:40 UT and reaches its maximum ($\sim$150 km s$^{-1}$) at $\sim$06:55:17 UT. After the velocity reaches its second peak ($\sim$140 km s$^{-1}$) at $\sim$06:56:05 UT, it gradually decreases to nearly zero at $\sim$07:00:05 UT. The variation behaviour at location B (the bottom panel of Figure \ref{fig6}) is similar to that of location A except for a slight difference in timing. The redshifts at location B firstly appear at $\sim$06:53:10 UT and rise up to the first peak ($\sim$150 km s$^{-1}$) at $\sim$06:54:30 UT. Like at location A, the Doppler velocity at location B shows two peaks in coincidence with two peaks in intensity (as indicated by the two vertical dash-dotted lines). By comparison, the Doppler velocity at location B reaches its first peak about one minute earlier than that at location A. The second peak of the velocity, however, appears at the same time for both locations A and B.

The time profiles of line intensity and Doppler velocity at locations C and D (see Figure \ref{fig7}) are distinct from those mentioned above. It is seen that, at location C, the intensity increases at $\sim$06:54:00 UT and shows some fluctuations. For the velocity, it starts to rise earlier at $\sim$06:52:10 UT and decreases to zero at $\sim$06:58:50 UT, also showing some fluctuations with amplitudes ranging from $\sim$30 to $\sim$50 km s$^{-1}$. Location D also reveals a fluctuation behaviour in the time profiles of intensity and velocity. Compared with the redshift velocities at locations A and B, the velocities at locations C and D are much smaller.

\section{Discussions}
\label{sec4}

As shown above, the two sets of locations exhibit some distinct spectral features that are supposed to originate from different processes. As the C1.7 flare studied here is small in size and especially complex in morphology, it is somewhat difficult to determine the precise positions of the four selected locations, i.e., whether at flare ribbons or on flare loops, from the present data. In this section, we only provide some possibilities or speculations for the physical origin of the redshifts and blueshifts observed at these locations.
 
For locations A and B, there show up continuum emission and cool narrow lines, which seem to support that they correspond to flare ribbons. However, such high-speed ($\sim$150 km s$^{-1}$) redshifts have scarcely been reported in the cool lines at flare ribbons and seem to be hard to explain using the ribbon scenario. In particular, relatively strong emission as well as evident redshifts are detected in the hot \ion{Fe}{21} line at these two locations, which most likely originate from flare loops \citep[e.g.,][]{Tian2014b,Young15, Tian2016,Polito2018}. In fact, some loop-like structures can be seen in the SJIs as well as AIA images at these two locations. Based on all of these, we conjecture that the flare loops probably overlap with the flare ribbons along the line of sight at locations A and B. {In the following, we consider several possibilities for the origin of the high-speed ($>$100 km s$^{-1}$) redshifts or downflows at these two locations presuming that they are mainly contributed by flare loops:} (1) project effect, (2) filament plasma draining, (3) hot plasma cooling down, and (4) reconnection outflows.

Firstly, considering that the flare under study is near the solar limb, we need to check the possible consequence of the projection effect. Sometimes, a particular viewing angle could attain redshifts for actual upflows. However, this is not the case here, since the angle between the line of sight and the loop axis seems to be still acute. 

\begin{figure*}
	\centering
	\includegraphics[width=\linewidth]{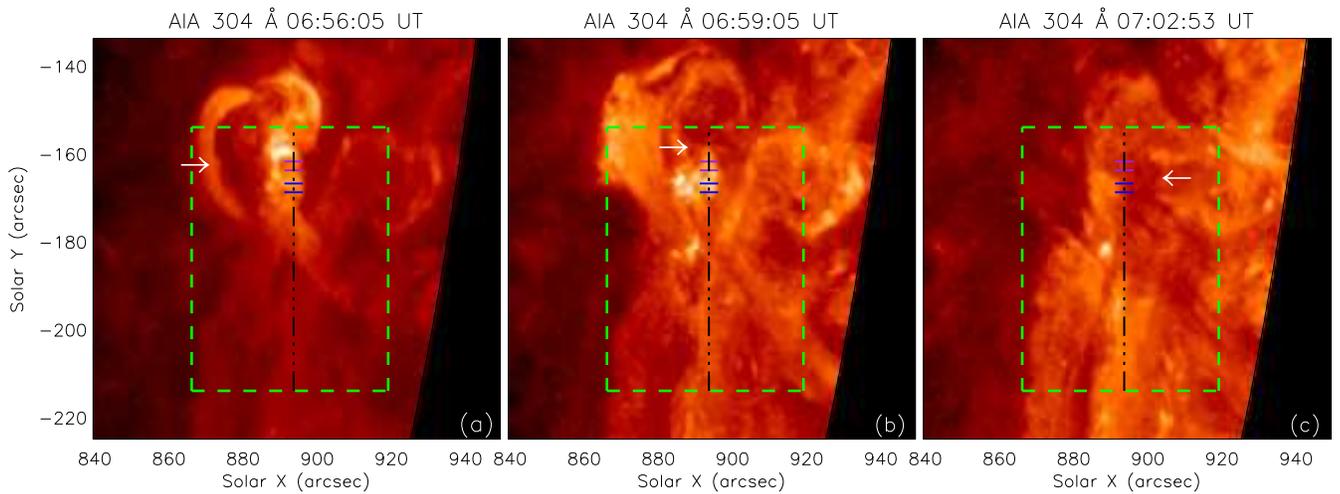}
	\caption{
AIA 304 \AA\ images showing the evolution of the filament draining process. In each panel,  {the purple and blue bars denote four locations that are selected for study.} 
The green dashed box refers to the FOV of the {\em IRIS} SJIs and the black dash-dotted line refers to the slit position.  {The white arrows indicate the filament plasma draining process.}}
    \label{fig8}
\end{figure*}

Secondly, we notice that there appears some filament plasma draining in this C1.7 flare as revealed by AIA 304 \AA\ images. However, after a careful check of the images, we find that the draining starts to appear in the FOV of {\em IRIS} at $\sim$06:56 UT, and then moves through the {\em IRIS} slit at $\sim$06:59 UT (Figure \ref{fig8}), when the high-speed redshifts at locations A and B have almost disappeared. Therefore, this filament plasma draining is unlikely responsible for the origin of the high-speed downflows.

Thirdly, when some hot plasma, say, the evaporation plasma, cools down, it will produce significant redshifts, particularly in the decay phase of the flare. However, we notice that the downflows mostly appear before the SXR emission peak time, i.e., in the rise phase of the flare. Hence the cooling plasma may not be the cause of the high-speed redshifts in the rise phase.
 
\begin{figure*}
	\centering
	\includegraphics[width=\linewidth]{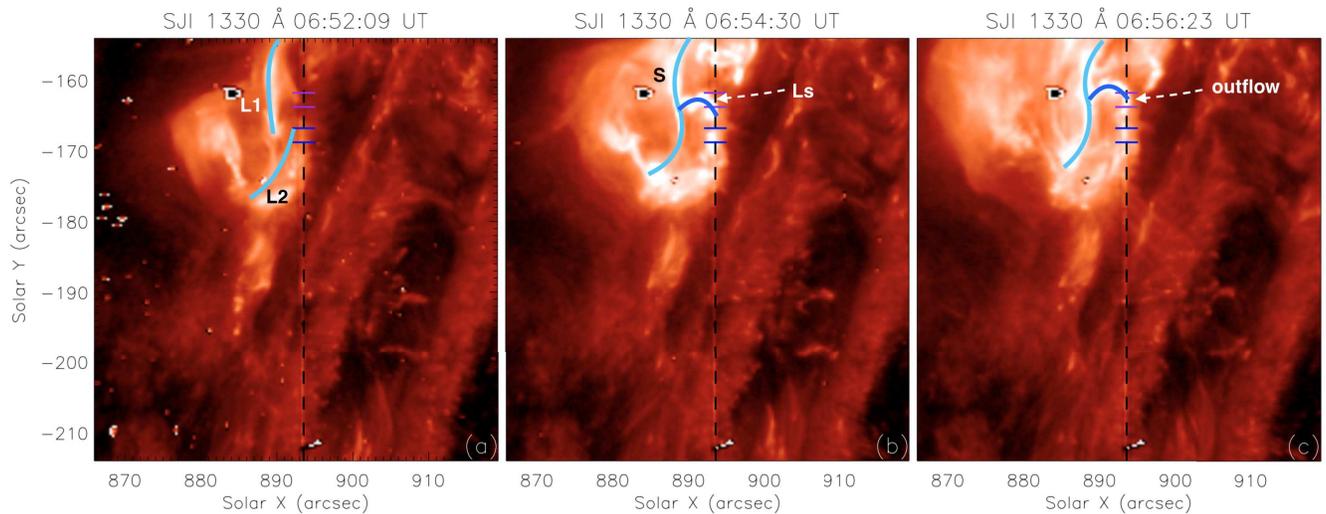}
	\caption{
SJIs 1330 \AA\ showing the evolution of the reconnection process. In each panel, the dashed line refers to the position of the {\em IRIS} slit, on which  {four bars mark four locations that are selected for study.}
The cyan lines in panel (a) delineate two coronal loops (L1 and L2) in the initial phase of the flare, while the blue line in panels (b) and (c) shows the small flare loop (Ls) that is formed during the magnetic reconnection. The cyan curve S in panels (b) and (c) shows the sigmoid structure after reconnection.}
    \label{fig9}
\end{figure*}

Finally, the remaining possibility is that the high-speed downflows are a result of magnetic reconnection. This is illustrated in Figure \ref{fig9}. At the initial time, two coronal loops L1 and L2 are observed around the {\em IRIS} slit (Figure \ref{fig9}{(a)}). A few minutes later ($\sim$06:54 UT), these two loops approach and magnetic reconnection occur between them, which produces a small flare loop Ls as well as the sigmoid structure S (Figures \ref{fig9}(b) and (c)). The reconnection heats the plasma and drives plasma outflows that move along the newly formed flare loop Ls. One of the outflows is located near the slit positions A and B while the other is out of the slit region. The sigmoid structure then loses its balance and undergoes a subsequent eruption in the corona. The illustration here is in accordance with the tether-cutting (TC) model proposed by \cite{Moore2001}. The high-speed redshifts are likely due to the outflow of the magnetic reconnection near the footpoint of Ls. Such high-speed Doppler velocities have also been reported by \cite{Chen2016}, in which, however, the reconnection is supposed to take place in the lower atmosphere due to magnetic cancellation. {Here it should also be mentioned that the accompanied redshifts in the hot \ion{Fe}{21} line might be caused by a retracting of hot flare loops \citep[e.g.,][]{Tian2014b} or termination shocks \citep[e.g.,][]{Polito2018,Shen2018}.}
 
{As regards locations C and D, they might correspond to flare loops and their behaviours could be explained by a loop scenario suitably. The gentle blueshifts in the hot \ion{Fe}{21} line as well as redshifts in the cool lines of \ion{Si}{4}, \ion{C}{2}, and \ion{Mg}{2} exhibit some fluctuations throughout the flare time, which are likely caused by hot plasma filling and cool plasma draining in the flare loops, respectively.}

\section{Conclusions}
\label{sec5}

In this paper, we have presented spatio-temporal variations of the intensity and Doppler velocity of several UV lines including \ion{Si}{4}, \ion{C}{2}, \ion{Mg}{2}, and \ion{Fe}{21} in a C1.7 flare observed by {\em IRIS}. Two sets of locations (A \& B and C \& D) are selected to reveal the typical features of the flare in detail. It is found that both sets of locations show evident brightenings but some distinct features in the line profiles, Doppler velocities, and their temporal evolutions. At the first set of locations A and B, the cool \ion{Si}{4}, \ion{C}{2}, and \ion{Mg}{2} lines exhibit significant redshifts with the velocities as high as 150 km s$^{-1}$. In the mean time, the hot \ion{Fe}{21} line shows redshifts with a velocity of a few km s$^{-1}$. The strong redshifts in the cool lines mainly show up in the rise phase of the flare, then increase rapidly, and proceed into the decay phase. By contrast, at the second set of locations C and D, the cool lines primarily show gentle redshifts with the velocities only up to tens of km s$^{-1}$. Simultaneously, the hot \ion{Fe}{21} line is blueshifted. The time profiles of line intensity and Doppler velocity at the second set of locations display some fluctuations throughout the flare period. All these distinct features suggest that different physical processes play a role at different flare regions. The high-speed redshifts in the cool lines, most likely originating from flare loops, is thought to be caused by magnetic reconnection outflows, {while the gentle redshifts in the cool lines could be regarded as a result of plasma draining in the flare loops.}

\acknowledgments
 {{\em IRIS} is a NASA small explorer mission developed and operated by LMSAL with mission operations executed at the NASA Ames Research center and major contributions to downlink communications funded by the Norwegian Space Center (NSC, Norway) through an ESA PRODEX contract. {\em SDO} is a mission of NASA’s Living With a Star Program. The authors thank the referee for the constructive comments that help to improve the manuscript  {substantially}.} This work was supported by NSFC under grants 11733003, 11873095, and 11903020, and NKBRSF under grant 2014CB744203. Y.L. is also supported by the CAS Pioneer Talents Program for Young Scientists and XDA15052200, XDA15320301 and XDA15320103. Y.-A.Z. would like to thank Yuhao Zhou for helpful discussions.

\clearpage

\clearpage
\end{document}